\documentclass{aankg}
\usepackage{graphics}
\newcommand{\doe}
{This work was supported by the
Director, Office of Energy Research,
Office of High Energy
and Nuclear Physics,
Division of Nuclear Physics,
of the U.S. Department of Energy under Contract
DE-AC03-76SF00098.}

\newcommand{\beqn}{\begin{eqnarray}}
\newcommand{\eeqn}{\end{eqnarray}}

\begin{document}

   \thesaurus{06     
              (19.53.1;  
               19.63.1)} 
   \title{Evolution from Canonical to Millisecond Pulsar
through the X-ray
Accretion Stage}


   \author{Norman K. Glendenning \and
   Fridolin Weber
          }


\institute{Nuclear Science Division  \&
   Institute for Nuclear and Particle Astrophysics\\
     Lawrence Berkeley  National Laboratory,
	  MS: 70A-319 \\ Berkeley, California 94720
	  }

 \date{\today}

\authorrunning{Glendenning \& Weber}
   \titlerunning{Evolution from Canonical to Millisecond Pulsar through
   the X-ray Accretion Stage}
   \maketitle

   \begin{abstract}
We model the evolution of canonical pulsars from the death line to 
millisecond pulsars through the X-ray neutron star
stage of accretion from a low-mass
companion. We trace this evolution in magnetic field strength
starting at $B=10^{12}$ to $10^8$ G and in period of about 1 s to
milliseconds. Important factors are accretion rate and the decay
rate of the magnetic field. A broad swathe is traced in the $B-P$
plane according to the value of these factors, which represent
different conditions of the binary pair. An important ingredient
is the dependence of the stellar moment of inertia on rotation frequency (time).

   \end{abstract}

%

\section{Background}

Pulsars are believed to have formed from the collapsed cores of massive stars, and are spun up by angular momentum conservation
to moderate frequencies of 30 Hz
 or so, like the Crab pulsar. 
 Thereafter, they radiate their angular momentum through
the strong magnetic dipole field which is fixed in the star, ever more slowly as the rotational rate decreases. Depending on the magnetic field strength and  frequency, the radiation mechanism turns off and canonical pulsars disappear from the radio sky at frequencies of 1 Hz or there about, depending 
on the magnetic field strength.
 They have crossed what has been referred to as the ``death line''. Millisecond pulsars are thought to originate from these radio silent stars, being
spun up by accretion from a low-mass companion. 
Their magnetic fields are typically $10^8$ to
$ 10^9$ Gauss, as compared to $10^{11}$ to $10^{13}$  Gauss for canonical pulsars. Either the field decays
ohmically during the long epoch in which they are radio and X-ray silent, or the field is partially destroyed by the accretion process, or by some
as yet unknown mechanism.
In this intermediate accretion phase, they are X-ray neutron          stars. The missing link between canonical
pulsars with mean period of 0.7 s and millisecond pulsars was recently
discovered, an X-ray star with period of 2.5 ms
(\cite{klis98:b,chakrabarty98:a,klis00:a})

So far as we know, the evolution from canonical to millisecond pulsar has never been modeled. We do so here. We find essentially a continuum of evolutionary tracks according to the rate at which matter is accreted from the
companion, and the rate at which the magnetic field decays.  The evolutionary tracks essentially fill all the space in the B-P plane, starting at the death line at fields typical of canonical pulsars,
and extending downward in field strength, broadening to fill the space on both sides of the deathline, and extending to the small periods of  millisecond pulsars.     All are potential tracks of some particular binary pair, since
accretion rates vary by several orders of magnitude and presumably so do decay rates of the magnetic field.

\section{Evolution from canonical to ms pulsars} 
There are two distinct aspects to developing an evolutionary framework.
One has to do with the accretion process itself, which has been developed by a number of authors in the framework of classical physics. We quote some of the important results below and refer to the 
original literature for details
(\cite{elsner77:a,ghosh77:a,lipunov92:book}).

The other aspect has to do with the structure of the neutron star and its response to added mass, but most especially to 
its response to changes in rotational frequency due to the changing 
centrifugal forces.
 Typically, the moment of inertia has been computed in general relativity for a non-rotating star (\cite{hartle67:a,hartle68:a}).
It is based on the Oppenheimer-Volkoff metric. However, for the purpose of tracing the evolution of an accreting star from $\sim 1$ Hz to 400 to 
600 Hz we do not
neglect the response of the shape, structure and composition of the star as it is spun up over this vast range of  frequencies from essentially zero to values that approach
the Kepler frequency. Nor do
we neglect the dragging of local inertial frames.
These features are included in our calculation of the
tracks of neutron stars from canonical objects starting with large fields and very small frequencies at the ``death line'' 
to the small fields but rapid rotation of millisecond pulsars.  
However, the expression
of the moment of inertia and a definition of the various factors that enter are too long to reproduce here. We refer instead to our derivation
given in Ref.\  (\cite{glen92:crust}). 

The spin-up torque of
the accreting matter causes a change in the star's angular momentum according
to the relation (\cite{elsner77:a,ghosh77:a,lipunov92:book})
\begin{eqnarray}
{{dJ} \over {d t}} = {\dot M} {\tilde l}(r_{\rm m}) - N(r_{\rm c}) \,
,
\label{eq:dJdt}
\end{eqnarray}
where $\dot{M}$ denotes the accretion rate and ($G=c=1$) 
\begin{eqnarray}
{\tilde l}(r_{\rm m}) = \sqrt{M r_{\rm m}} 
\label{eq:l}
\end{eqnarray}
is the angular momentum added to the star per unit mass of accreted
matter. The quantity $N$ stands for the magnetic plus viscous torque
term,
\begin{equation}
N(r_{\rm c}) = \kappa \, \mu^2 \, r_{\rm c}^{-3} \, ,
\label{eq:N}
\end{equation}
with $\mu \equiv R^3 B$ the star's magnetic moment ($\kappa\sim 0.1$).
We assume that the magnetic field evolves according to
\begin{equation}
B(t) = B(\infty) + (B(0) - B(\infty)) e^{-t/t_{\rm d}}
\label{bevolution}
\end{equation} 
where  $B(\infty)=10^8$ G,
$B(0)=10^{12}$ G and $t_{\rm d} = 10^5$ 
to $10^7$ yr (cf. \cite{konar}).  
However, we shall also make a comparison with a
purely exponential decay. The
quantities $r_{\rm m}$ and $r_{\rm c}$ denote the radius of the inner edge of
the accretion disk and the co-rotating radius, respectively, and are given by
$(\xi \sim 1)$
\begin{equation}
  r_{\rm m} = \xi \, r_{\rm A} \, ,
\label{eq:r_m}
\end{equation}
and 
\begin{eqnarray}
r_{\rm c} = \left( M \Omega^{-2} \right)^{1/3} \, .
\label{eq:r_c}
\end{eqnarray} Accretion will be inhibited by a centrifugal barrier if the
neutron star's magnetosphere rotates faster than the Kepler frequency at the
magnetosphere. Hence $r_{\rm m} < r_{\rm c}$, otherwise accretion onto the star
will cease.  The Alf\'en radius $r_{\rm A}$, where the magnetic energy density
equals the  kinetic energy density
of the accreting matter, in Eq.\ (\ref{eq:r_m})
is defined by
\begin{equation}
r_{\rm A} = \left( { {\mu^4} \over {2 M \dot{M}^2} } \right)^{1/7} \, .
\label{eq:r_A}
\end{equation}
The rate of change of a star's angular frequency $\Omega$ $(\equiv
J/I)$ then follows from Eq.\ (\ref{eq:dJdt}) as
\begin{equation}
  I(t) {{d\Omega(t)} \over {d t}} = {\dot M} {\tilde l}(t) - \Omega(t)
  {{dI(t)}\over{dt}} - \kappa \, \mu(t)^2 \, r_{\rm c}(t)^{-3} \, ,
\label{eq:dOdt.1}\label{spinevolution}
\end{equation} with the explicit time dependences as indicated.  Evidently, the
second term on the right-hand-side of Eq.\ (\ref{eq:dOdt.1}) depends linearly
on $\Omega$ to leading order, while the third terms grows quadratically with
$\Omega$. As mentioned, the change of the moment of inertia with time
(or correspondingly frequency) is an essential part of the evolution and we 
do not neglect this term in the frequency evolution equation.

The stellar  model from which the moment of inertia is computed
 is described as follows.
Neutron star matter has   a charge
neutral  composition of hadrons consisting of members of the
baryon octet together with leptons.
The properties of such matter are calculated in a covariant mean field
theory as described in Refs.\
(\cite{glen85:b,kapusta90:a,kapusta91:a,mybook}).
The values of the
parameters that define the coupling constants of the theory  are
certain fairly well constrained properties of nuclear matter
and hypernuclei as described in the references;
(binding energy of symmetric nuclear matter $B/A=-16.3$ MeV,
saturation density $\rho=0.153 {\rm ~fm^{-3}}$, compression modulus
$K=300$ MeV, symmetry energy coefficient $a_{{\rm sym}}=32.5$ MeV,
nucleon effective mass at saturation $m^{\star}_{{\rm sat}}=0.7m$ and
ratio of hyperon to nucleon couplings $x_\sigma=0.6,~x_\omega=0.653=x_\rho$
that yield, together with the foregoing parameters, the correct
$\Lambda$ binding in nuclear matter (\cite{glen91:c})). The above are
representative, but by no means well determined in all cases. But the
effect of 
the response of the moment of inertia to frequency  on the time 
evolution of the accreting system should be approximately represented.

\section{Evolutionary Tracks in the $B-P$ Plane}

\begin{figure}[htb]
\begin{center}
\resizebox{7cm}{!}{\includegraphics{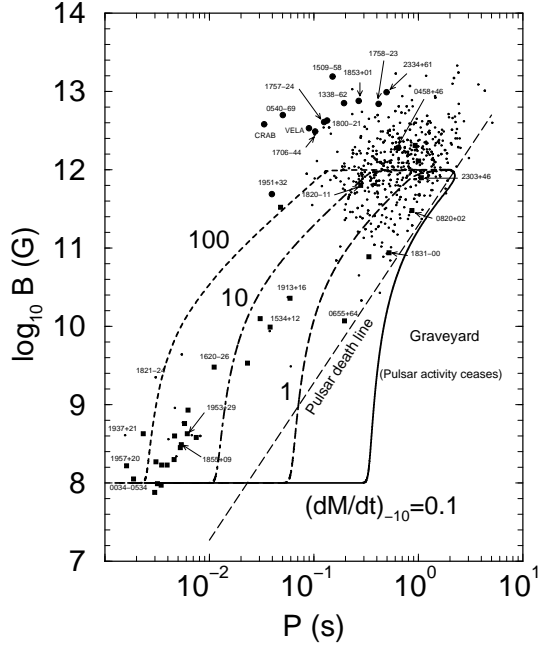}}
{ \caption { \label{bp_4} Evolutionary tracks traced by neutron stars
in the X-ray accretion stage, 
beginning 
on the death line with large $B$ field
and ending as  millisecond stars, for various accretion rates.
Here $t_{{\rm d}}=10^6$yr.
}}
\end{center}
\end{figure}

As a first orientation as to our results and how they relate to 
known pulsars as regards their magnetic field strength and their
rotational period, we show the evolutionary tracks for four different
accretion rates given in units of $\dot{M}=10^{-10}$ solar masses per year in 
Fig.\ \ref{bp_4}.
The decay rate of the magnetic field is taken to have the value
$ t_{{\rm d}}=10^6$ yr in each case. The initial condition is that
after some indefinite era on the death line at a field strength of
$10^{12}$ Gauss, accretion from a companion commences. The X-ray neutron star gains angular momentum and its period decreases and over a longer 
timescale, the magnetic field decays. One can see already that a wide 
swathe of $B$ and $P$ is traced out.

\begin{figure}[htb]
\begin{center}
\resizebox{7cm}{!}{\includegraphics{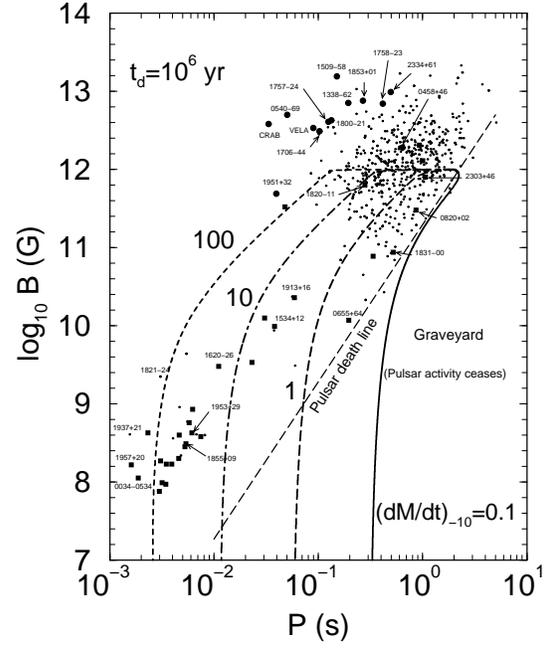}}
{ \caption { \label{bp_4e} Evolutionary tracks traced by neutron stars
in the X-ray accretion stage,
beginning
on the death line with purely exponential decay of the $B$ field.
As in Fig.\ \protect\ref{bp_4}, $t_{{\rm td}}=10^6$yr.
}}
\end{center}
\end{figure}

In the above example, the field was assumed to decay to a finite
asymptotic value of $10^8$ Gauss.
 A very different assumption, namely that the field
decays eventually to zero, $B(t)=B(0)e^{-t/t_{{\rm d}}}$, modifies
only the results below the asymptotic value, as is seen by comparing
Fig.\ \ref{bp_4} and \ref{bp_4e}. 
However, the conclusion concerning the origin of millisecond pulsars is 
quite different. For purely exponential decay, one would conclude that
high frequency pulsars are created only in high accretion rate
binaries.

\begin{figure}[htb]
\begin{center}
\resizebox{6cm}{!}{\includegraphics{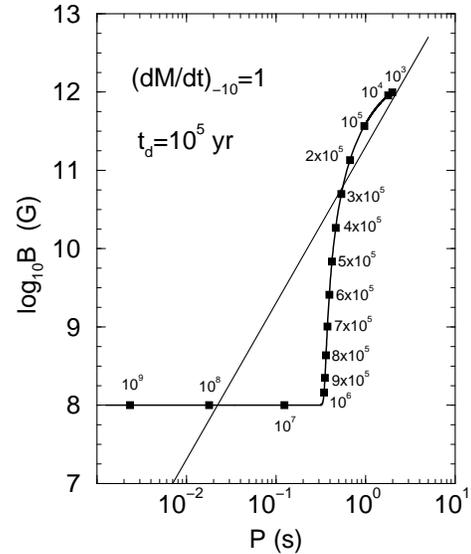}}
{ \caption { \label{time_tag} Time tags are shown for one of the
evolutionary tracks.
}}
\end{center}
\end{figure}

In the remainder of the paper, we assume the field decays to an asymptotic 
value, since from the above comparison we see how exponential decay would
modify the picture.
We show time tags on a sample track in Fig.\ \ref{time_tag} which provides
some sense of time lapse. The first part of a track is traversed in short
time, but the remainder ever more slowly. This shows up also in
$dP/dt$ as a function of time.

\begin{figure}[htb]
\begin{center}
\resizebox{6cm}{!}{\includegraphics{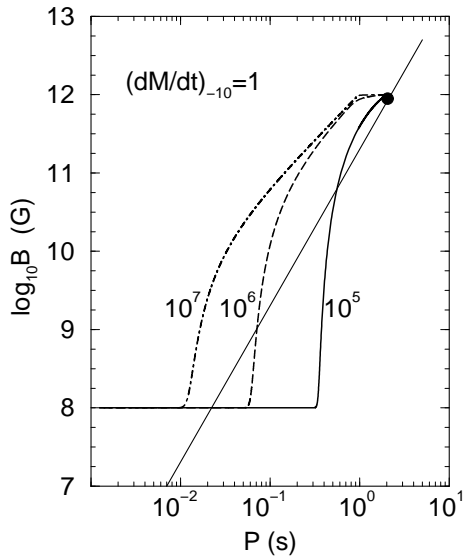}}
{ \caption { \label{bp_a} Evolutionary tracks for a neutron star
starting at the death line and evolving by accretion to lower
field and high frequency for an accretion rate 1 in units of 
$10^{-10}$ solar masses per year, and for three values of the
magnetic field decay rate $t_{{\rm d}}$ as marked.
}}
\end{center}
\end{figure}

\begin{figure}[htb]
\begin{center}
\resizebox{6cm}{!}{\includegraphics{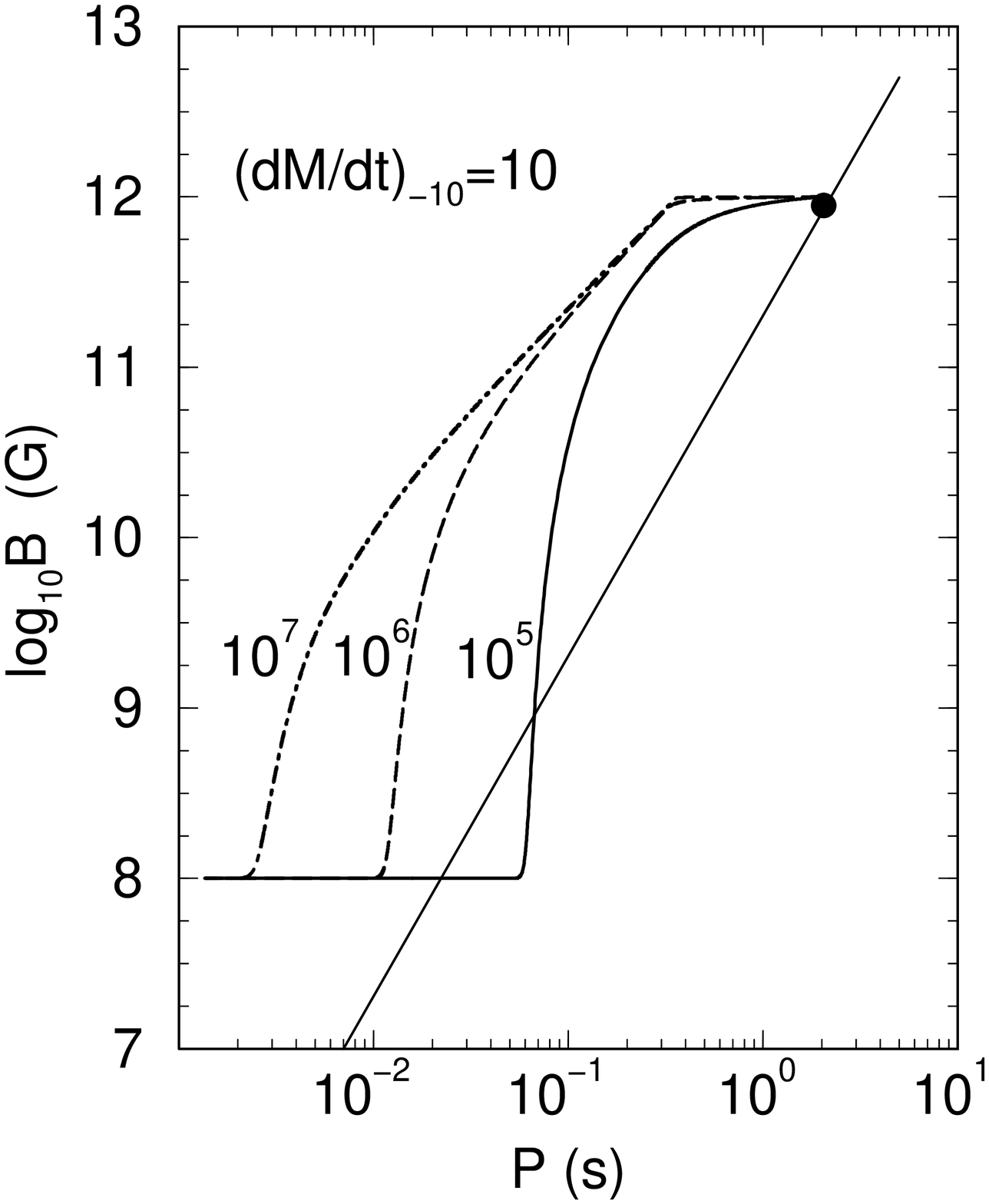}}
{ \caption { \label{bp_b} Similar to Fig.\ \protect\ref{bp_a}
but with a different accretion rate as marked.}}
\end{center}
\end{figure}

\begin{figure}[htb]
\begin{center}
\resizebox{6cm}{!}{\includegraphics{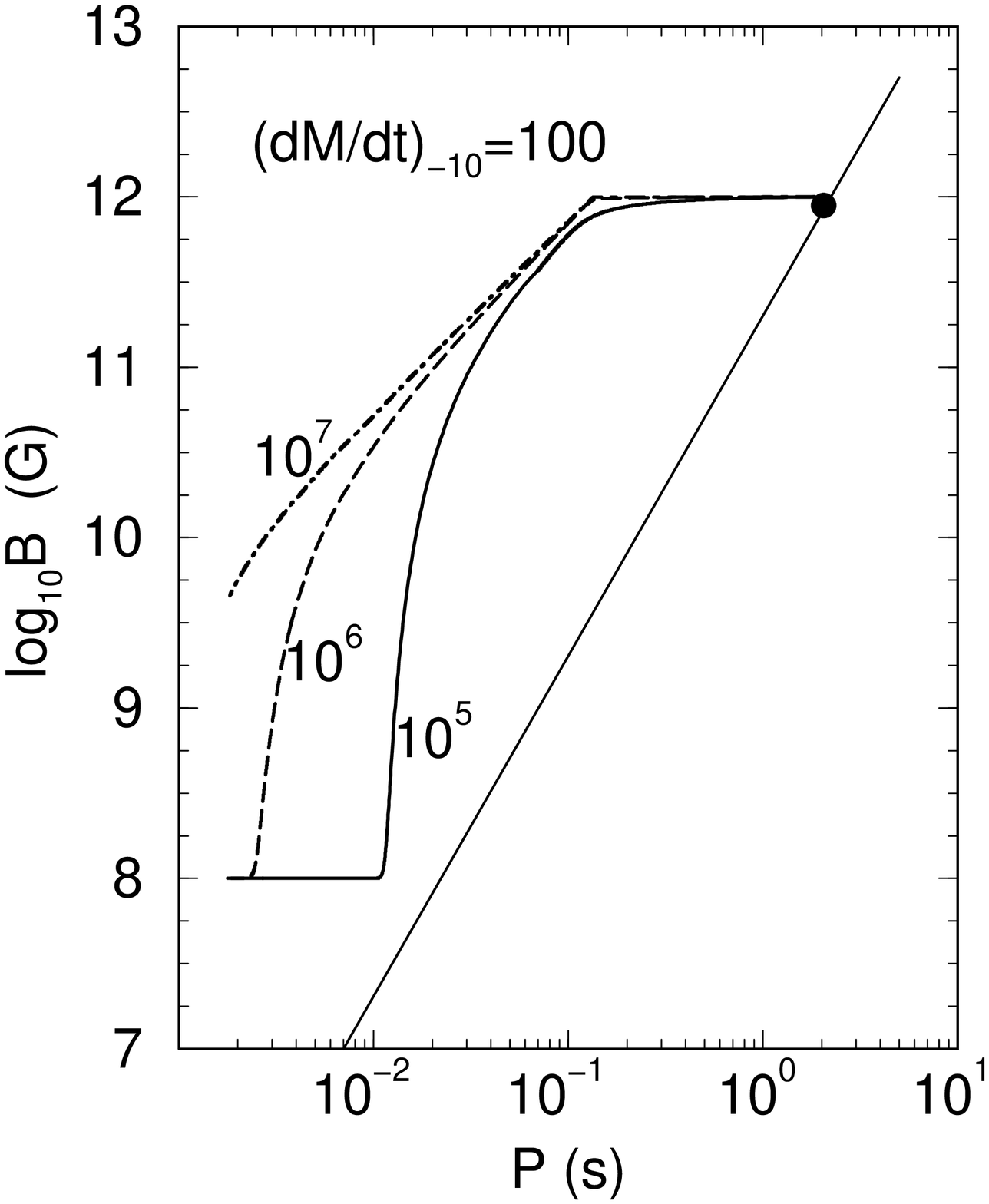}}
{ \caption { \label{bp_c} Similar to Fig.\ \protect\ref{bp_a}
but with a different accretion rate as marked.}}
\end{center}
\end{figure}

For each of three accretion rates we show the dependence on three 
field decay rates in Figs.\ \ref{bp_a}, \ref{bp_b} and \ref{bp_c}.
Depending on decay rate of the field and accretion rate, an X-ray
neutron star may spend some time on either side of the death line,
but if it accreted long enough, always ends up as a candidate for
a millisecond pulsar {\sl if} the magnetic filed decays to an
assymptotic value such as was assumed. However, if the field decays
exponentially to zero, only high accretion rates would lead to 
millisecond pulsars. Of course, if accretion turns off at some time, 
the evolution is arrested. 

\section{Summary} We have computed the evolutionary tracks 
in the $B-P$ plane
due to mass accretion onto 
neutron stars beginning at the death line with a typical field 
strength of $10^{12}$ Gauss, to shorter periods and low
fields. 
According to the assumed accretion rate and decay
constant for the magnetic field, the tracks indicate that the
individual binaries with characteristics ranging from Z to Atoll
sources will evolve along paths that cover a broad swathe
in the $B-P$ plane. These
 include tracks of X-ray stars
corresponding to low accretion rates
 that follow a path beyond the
death line in the so-called ``graveyard'.

We have assumed two  particular forms for the law of decay of the
magnetic field. (1) The field  approaches an asymptotic value of $10^8$ Gauss
such as is typical of millisecond pulsars. This assumption leads to a 
particular form for the
termination of
evolutionary tracks. All accreters, no matter what the accretion rate, 
will end with millisecond periods, unless accretion ceases beforehand. 
(2) If instead, we had assumed a purely exponential decay, 
the tracks would not tend to an asymptotic value, 
but would continue to decrease in the strength of $B$. The tracks would 
still cover a broad swathe in the $B-P$ plane.  But one would conclude that 
only the higher accretion rate binaries, particularly
the Z-sources, could produce
millisecond period neutron stars. If acccretion continues for too long a time,
the neutron star will be carried to very low fields and across the death
line, or an overcritical mass will have been accreted, leading instead to
a black hole.

\begin{acknowledgements}
		  \doe ~F. W. was supported by a grant from
		    the Deutsche Forschungsgemeinschaft (DFG).
		  \end{acknowledgements}


\end{document}